\documentclass[iop,apj,twocolappendix,numberedappendix,appendixfloats]{emulateapj}
\pdfoutput=1 

\usepackage[breaklinks,colorlinks,citecolor=cyan, linkcolor=magenta]{hyperref} 
\usepackage{apjfonts} 
\usepackage{amsmath}
\usepackage{relsize}
\usepackage{amsbsy}
\usepackage{natbib} 
\usepackage{enumitem} 
\usepackage{mathtools}
\DeclarePairedDelimiter{\abs}{\lvert}{\rvert}

\shortauthors{Behmard et al.}

\begin{document}

\received{}
\accepted{}


\title{Data-Driven Spectroscopy of Cool Stars at High Spectral Resolution}



\author{Aida Behmard\altaffilmark{1}}
\affil{$^{1}$Division of Geological and Planetary Sciences, California Institute of Technology, Pasadena, CA 91125, USA}

\author{Erik A. Petigura\altaffilmark{2,3}, Andrew W. Howard\altaffilmark{2}}
\affil{$^{2}$Cahill Center for Astrophysics, California Institute of Technology, Pasadena, CA 91125, USA}
\affil{$^{3}$Hubble Fellow}




\begin{abstract}

The advent of large-scale spectroscopic surveys underscores the need to develop robust techniques for determining stellar properties ("labels", i.e., physical parameters and elemental abundances). However, traditional spectroscopic methods that utilize stellar models struggle to reproduce cool ($<$4700 K) stellar atmospheres due to an abundance of unconstrained molecular transitions, making modeling via synthetic spectral libraries difficult. Because small, cool stars such as K and M dwarfs are both common and good targets for finding small, cool planets, establishing precise spectral modeling techniques for these stars is of high priority. To address this, we apply \emph{The Cannon}, a data-driven method of determining stellar labels, to Keck High Resolution Echelle Spectrometer (HIRES) spectra of 141 cool ($<$5200 K) stars from the California Planet Search. Our implementation is capable of predicting labels for small ($<$1 $R_{\odot}$) stars of spectral types K and later with accuracies of 68 K in effective temperature ($T_{\textrm{eff}}$), 5\% in stellar radius ($R_{*}$), and 0.08 dex in bulk metallicity ([Fe/H]), and maintains this performance at low spectral resolutions ($R$ $<$ 5000). As M-dwarfs are the focus of many future planet-detection surveys, this work can aid efforts to better characterize the cool star population and uncover correlations between cool star abundances and planet occurrence for constraining planet formation theories.

\end{abstract}


\maketitle



\section{Introduction} \label{sec:intro}
Precise determination of stellar properties (e.g., masses, radii, effective temperatures, elemental abundances) is a challenging, yet essential component of stellar and planetary astrophysics. Accurate measurements of masses ($M_{*}$), radii ($R_{*}$), and temperatures ($T_{\textrm{eff}}$) are crucial for vetting models of stellar structure and evolution, and the chemical compositions of stellar photospheres reflect formation histories and can link stars to their parent molecular clouds, providing a window into galactic chemical evolution. The burgeoning field of exoplanets also calls for robust methods of determining stellar properties as characterization of planets is predicated on thorough characterization of their stellar hosts. 

Stellar spectroscopy has a rich history, beginning with Annie Jump Cannon and her colleagues at Harvard College Observatory who developed the current stellar classification system based upon visual inspection of spectral features. Modern spectroscopic methods involve matching information-rich portions of empirical spectra to benchmark or synthetic spectra generated from model stellar photospheres. Two commonly-used spectral modeling tools are \texttt{SME} and \texttt{MOOG} \citep{valenti96, sneden73}, both of which have undergone significant evolution since their inception (e.g., \citealt{valenti05, valenti09, deen13, brewer15, piskunov17}). However, current model photospheres are limited by an incomplete knowledge of the physics behind stellar attributes; they suffer from poorly constrained atomic and molecular opacities, often assume local thermodynamic equilibrium (LTE), and inaccurately model dynamical effects such as convection or stellar winds, if at all. While three-dimensional hydrodynamic models have been created that allow for non-LTE conditions, they still suffer from the other aforementioned drawbacks and are computationally expensive. Laboratory studies have refined atomic and molecular data and improved line lists, but departures from solar type atmospheres still present significant modeling challenges. 

Stars of spectral types K4 ($T_{\textrm{eff}}$ $\lesssim$ 4700 K) and later are particularly difficult to model with synthetic spectral techniques as their optical and NIR spectra feature dense clusters of molecular lines that lack reliable opacity data. In the optical regions of K and M dwarf spectra, TiO and VO bands are prominent, as well as hydride bands such as MgH, CaH, and FeH. The NIR regions of M dwarf spectra often feature H$_{2}$O (e.g., \citealt{rojas_ayala12}). Characterization of late-type stars such as M dwarfs is important because they are common, representing $\sim$75\% of stars in the solar neighborhood \citep{henry06}. Small, cool stars are also popular targets for exoplanet surveys as their low $M_{*}$ and $R_{*}$ result in deeper transit signals and larger Doppler shifts, increasing the probability of detecting and characterizing small planets. 

Empirical methods offer alternative routes for predicting K and M dwarf parameters and abundances. Common proper motion pairs of M dwarfs and F-, G-, K-type stars of known metallicities ([Fe/H]) can be used to calibrate M dwarf metallicities with equivalent widths (EWs) of NIR spectral features \citep{newton14, mann14}. Similarly, temperatures ($T_{\textrm{eff}}$) and stellar radii ($R_{*}$) can be calibrated with EWs of K and M dwarf NIR spectra \citep{newton15}, and parallaxes can provide further constraints on stellar properties \citep{mann15, mann17}. Empirical as opposed to synthetic spectral libraries composed of touchstone stars with well-measured properties are also capable of predicting accurate parameters for stars of mid-K spectral types and later \citep{yee17}. 

Another promising method for modeling cool stars is offered by \emph{The Cannon}, a data-driven approach to modeling spectroscopic data \citep{ness15}. In brief, \emph{The Cannon} predicts stellar parameters and elemental abundances from spectroscopic data via a two-step process: a "training step" where the spectra for a set of reference objects with well-determined parameters and/or abundances are used to construct a predictive model of the flux, and a "test step" where the model is used to infer those of objects given their spectra. Unlike traditional spectroscopic modeling methods, \emph{The Cannon} makes no use of physical stellar models, and does not require an accompanying library of synthetic spectra for reference. Here, we modify \emph{The Cannon} to optimize parameter and elemental abundance predictions for K and M dwarfs with HIRES spectra. 

Throughout this work, we refer to stellar parameters \emph{and} elemental abundances ($T_{\textrm{eff}}$, $R_{*}$, and [Fe/H]) as "labels" to be consistent with previous literature on \emph{The Cannon} (e.g., \citealt{ness15, casey16, ho17a}) and to adhere to machine learning/supervised methods terminology. We evaluate \emph{The Cannon}'s ability to predict stellar labels in our cool star sample with cross-validation experiments. Cross-validation was carried out by dividing a reference set of cool stars with well-determined labels into training and validation sets. The reference set is pulled from a library compiled by \citet{yee17} (see Section \ref{sec:sample} for more details). Performance was evaluated by examining how well \emph{Cannon}-predicted labels for the validation set matched those reported in the library. In Section \ref{sec:cannon}, we present \emph{The Cannon}, and outline our implementation and its performance on our cool star sample in Section \ref{sec:cvalidation}. We find that \emph{The Cannon} can predict labels with precisions of 68 K in $T_{\textrm{eff}}$, 5\% in R$_{*}$, and 0.08 in [Fe/H] (dex). Discussion of the results is presented in Section \ref{sec:discussion}.

\section{Cool Star Sample} \label{sec:sample}
Our spectral library was compiled by \citet{yee17} and consists of 404 touchstone stars originating from several source catalogs that span the spectral types $\sim$M5--F1 ($T_{\textrm{eff}}$ $\approx$ 3000--7000 K, $R_{*}$ $\approx$ 0.1--16 $R_{\odot}$). The stars have spectra obtained from HIRES at the Keck-I 10-m telescope \citep{vogt94} as part of the California Planet Search (CPS). For more details on CPS, see \citet{howard10}. The HIRES spectra are high-resolution ($R$ $\approx$ 60,000) and high signal-to-noise (S/N $>$ 40/pixel, with $\sim$80\% having S/N $>$ 100/pixel). The spectra originate from the middle HIRES detector CCD chip and contain 16 spectral orders. The HIRES blaze function has been removed and the spectra registered onto a common wavelength scale ($\lambda$ = 4990--6410 {{\AA}}) uniform in $\Delta$log$\lambda$ to ensure that linear velocity shifts correspond to linear pixel shifts \citep{yee17}. We confined the wavelength range to 13 orders ($\lambda$ = 4990--6095 {{\AA}}) to avoid redder portions of the middle HIRES CCD chip that are more affected by tellurics.

\begin{figure}[t]
		\centering
		\includegraphics[width=0.48\textwidth]{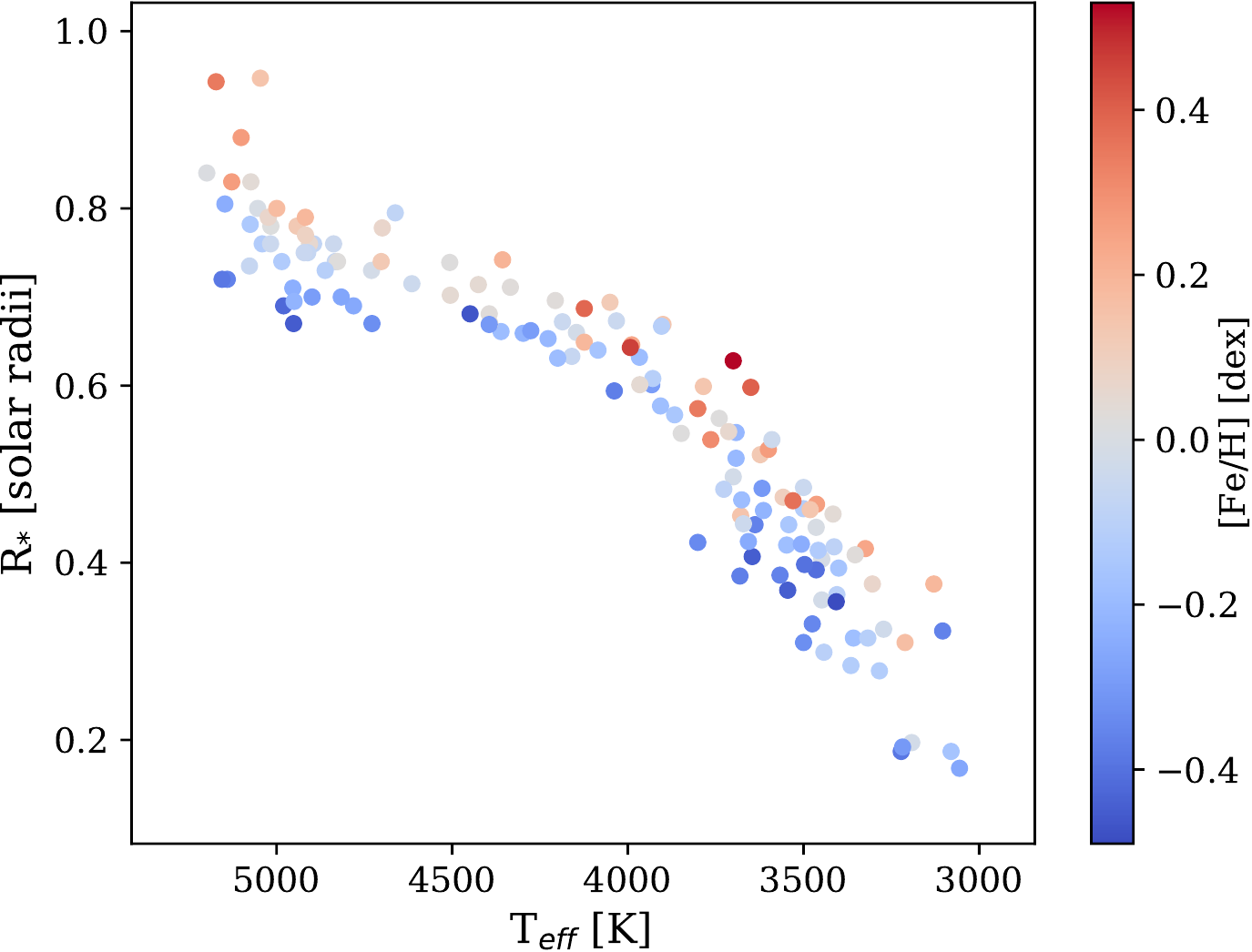}
		\caption{The domain of $T_{\textrm{eff}}$, $R_{*}$, and [Fe/H] for our reference sample of 141 cool stars pulled from the library outlined in \citet{yee17}. The cool stars have temperatures and radii that satisfy $T_{\textrm{eff}}$ $<$ 5200 K and $R_{*}$ $<$ 1 $R_{\odot}$.}
		\label{fig:figure1}
\end{figure} 

To isolate a cool star sample composed of K and M dwarfs, we employed radius and temperature cuts of $T_{\textrm{eff}}$ $<$5200 K and $R_{*}$ $<$ 1 $R_{\odot}$, leaving 141 stars. These cool stars are primarily drawn from the catalog described in \citet{mann15} with $T_{\textrm{eff}}$, $R_{*}$, and [Fe/H] determined from a combination of spectrophotometry, SED modeling, \emph{Gaia} parallaxes, and EW empirical relations (quoted uncertainties of 60 K, 3.8\%, and 0.08 dex, respectively). A smaller subset originate from the catalog compiled by \citet{vonbraun14}, and have interferometrically-determined $R_{*}$ (quoted uncertainties of $<$5\%). Many of the early K dwarfs in the sample have $T_{\textrm{eff}}$ and [Fe/H] determined from LTE spectral synthesis carried out by \citet{brewer16} with \texttt{SME} (quoted uncertainties of 60 K and 0.05 dex, respectively), while the sample mid to late K dwarfs have $T_{\textrm{eff}}$, $R_{*}$, and [Fe/H] determined from a combination of spectrophotometry, SED modeling, parallaxes, and \texttt{SME} analysis carried out by \citet{yee17} (quoted uncertainties of 5\%, 7.4\%, and 0.1 dex, respectively). Because most of these catalogs do not provide a complete set of $T_{\textrm{eff}}$, $R_{*}$, and [Fe/H] values, \citet{yee17} conducted an isochrone analysis using Dartmouth stellar models \citep{dotter08} to obtain a homogeneous label set, and took uncertainties as the 5th and 95th percentiles of the MCMC distributions that resulted from fitting to the stellar model grids. For more details on any of the library catalogs or the isochrone analysis procedure, see \citet{yee17}.

\section{The Cannon} \label{sec:cannon}
\subsection{Preparing HIRES spectra for The Cannon} \label{sec:prep}
\begin{figure*}[t]
		\centering
		\includegraphics[width=0.99\textwidth]{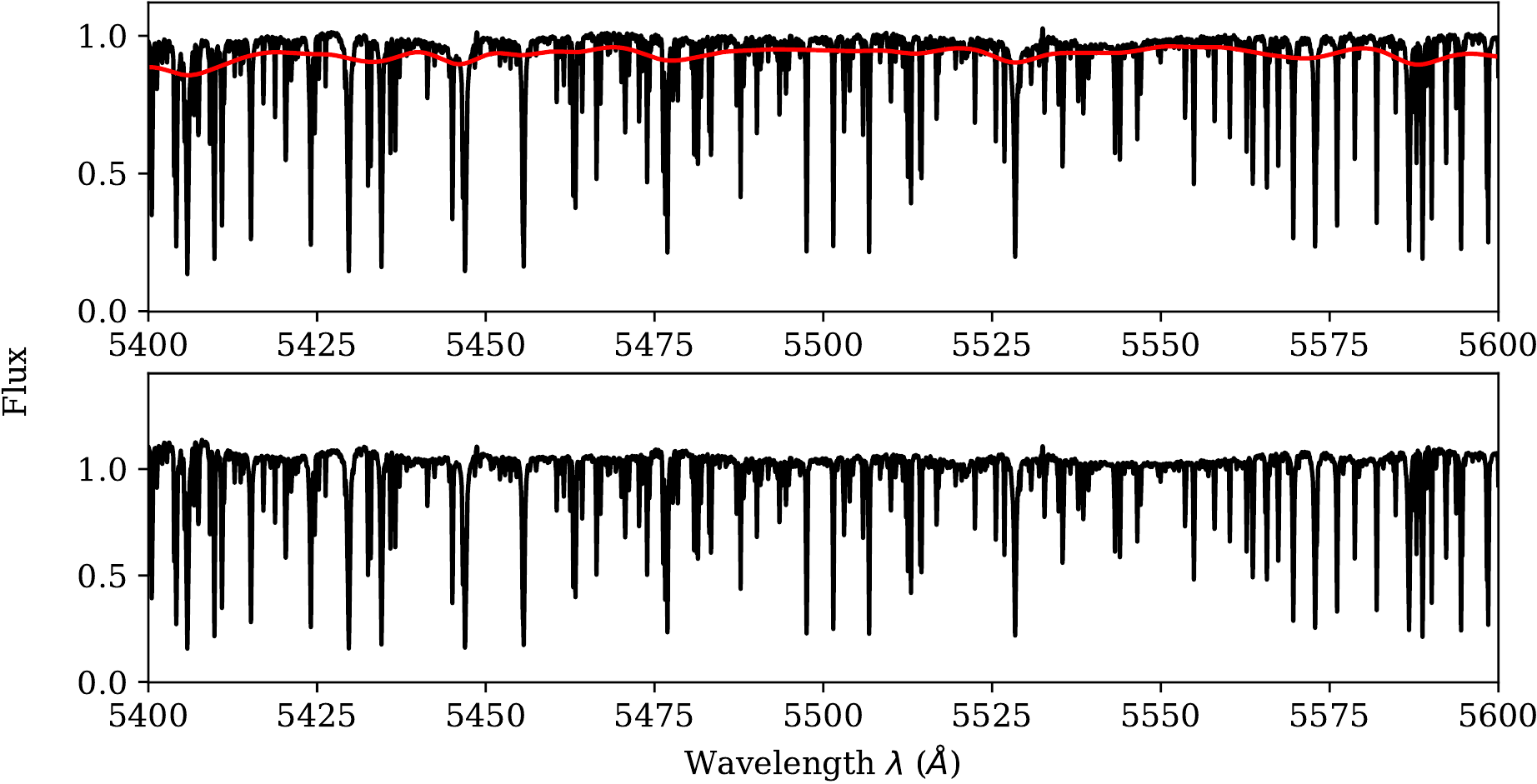}
		\caption{HIRES spectrum of a reference sample star (HD100623) before and after normalization. The top panel shows the pre-normalized spectrum overlaid with the Gaussian-smoothed version of itself in red, while the bottom panel shows the normalized spectrum after the Gaussian-smoothed signal was divided out. The displayed wavelength region ($\lambda$ = 5400-5600 {{\AA}}) is a subset of the full wavelength range and was chosen for better visualization of the spectrum and accompanying Gaussian-smoothed curve.}
		\label{fig:figure2}
\end{figure*}

To prepare the spectral library for \emph{The Cannon}, we must ensure that the spectra satisfy certain conditions; the spectra must share a common wavelength grid, be shifted onto the rest wavelength frame, share a common line-spread function, and be continuum-normalized via a method independent of S/N \citep{ness15}. The first two conditions are already satisfied for the library spectra, and we can assume that they effectively share a line-spread function, though there may be negligible variation due to variable observation seeing conditions. To carry out normalization, we applied error-weighted, broad Gaussian smoothing with

\begin{eqnarray}
\bar{f}(\lambda_{0}) = \frac{\sum_{j} (f_{j} \sigma_{j}^{-2}w_{j}(\lambda_{0}))}{\sum_{j} (\sigma_{j}^{-2} w_{j}(\lambda_{0}))} \hspace{0.5mm},
\end{eqnarray}
 
\noindent where $f_{j}$ is the flux at pixel $j$ of the wavelength range, $\sigma_{j}$ is the uncertainty at pixel $j$, and the weight $w_{j} (\lambda_{0})$ is drawn from a Gaussian:

\begin{eqnarray}
w_{j}(\lambda_{0}) = e^{-\frac{(\lambda_{0} - \lambda_{j})^{2}}{L^{2}}} \hspace{0.5mm} ,
\end{eqnarray}

\noindent where $L$ was chosen to be 3 {{\AA}}. If larger $L$ values are chosen for HIRES spectra, continuum-normalization begins to remove high resolution features. For reference, \citet{ho17a} used a width of $L$ = 50 {{\AA}} to normalize low-resolution Large Sky Area Multi-Object Fibre Spectroscopic Telescope (LAMOST) spectra ($R$ $\approx$ 1800). The Gaussian smoothing procedure is illustrated in Fig. \ref{fig:figure2}. 

\subsection{Training Step}
We used \emph{The Cannon 2}, the second implementation of \emph{The Cannon} developed by \citet{casey16}. Hereafter, we will refer to \emph{The Cannon 2} simply as \emph{The Cannon}. This version builds upon the original with additional features that are designed to aid prediction of a larger label set including elemental abundances that go beyond bulk metallicity ([Fe/H]), such as regularization. 

As outlined in Section \ref{sec:intro}, in the training step, \emph{The Cannon} uses a set of reference objects with well-determined labels to construct a predictive model of the flux at every pixel of the wavelength range that is a function of the stellar labels. Model construction is based on two assumptions: that continuum-normalized spectra with identical labels look identical at every pixel, and that the flux at every pixel in a spectrum changes continuously as a function of the stellar labels. While \emph{The Cannon} can be trained on any set of empirical spectra and their labels, the resultant model will only be capable of predicting labels for spectra with properties that are represented in the training set. In other words, \emph{The Cannon} is not able to accurately extrapolate outside the training set parameter space, so the training set spectra must be representative of the test set spectra in order to predict accurate label values. It is also important to note that the \emph{Cannon}-predicted labels will only be as accurate as those of the training set. 

The flux model $f_{j n}$ for a spectrum $n$ at pixel $j$ can be written as

\begin{eqnarray}
f_{jn} = \pmb{v}(l_{n}) \cdot \pmb{\theta}_{j} + e_{j n} \hspace{1mm},
\end{eqnarray}

\noindent where $\pmb{\theta}_{j}$ is the set of spectral model coefficients at each pixel $j$ and $\pmb{v}(l_{n})$ is a function of the label list $l_{n}$ that is unique for each spectrum $n$. The function $\pmb{v}(l_{n})$ is referred to as the "vectorizer" which can accommodate functions that are linear in the coefficients $\pmb{\theta}_{j}$, but not necessarily simple polynomial expansions of the label list $l_{n}$. The noise term is described by $e_{jn}$ and can be taken as sampled from a Gaussian with zero mean and variance $\sigma_{jn}^{2} + s_{j}^{2}$ where $\sigma_{jn}^{2}$ is the uncertainty reported on the input HIRES spectra (flux variance) and $s_{j}^{2}$ is the intrinsic scatter of the model at each pixel $j$. This intrinsic scatter can be likened to the expected deviation of the model from the spectrum at $j$. 

To determine the optimal model labels ($\pmb{\theta}_{j}$,$s_{j}^{2}$), we can relate the flux model to a single-pixel log-likelihood function:

\begin{eqnarray} \label{eq:eq4}
\textrm{ln}p(f_{jn} | \pmb{\theta}_{j},\pmb{v}(l_{n}),s_{j}^{2}) = - \hspace{0.5mm} \frac{[f_{jn}  \hspace{0.5mm} -  \hspace{0.5mm} \pmb{v}(l_{n}) \cdot \pmb{\theta}]^{2}}{\sigma_{jn}^{2} + s_{j}^{2}}  \hspace{0.5mm} -  \hspace{0.5mm}  \nonumber \\ \textrm{ln}(\sigma_{jn}^{2} + s_{j}^{2}) - \Lambda Q(\pmb{\theta}) \hspace{0.5mm},
\end{eqnarray}

\noindent where $\Lambda$ is a regularization parameter and $Q(\pmb{\theta})$ is a regularizing function that encourages the model coefficients $\pmb{\theta}_{j}$ to take on zero values, resulting in a simpler model that is less prone to overfitting. In the case of L1 regularization implemented within \emph{The Cannon}, the regularizing function takes the form

\begin{eqnarray}
Q(\pmb{\theta}) = \sum_{j=0}^{J-1} \abs{\theta_{j}}\hspace{0.5mm}.
\end{eqnarray}

L1 regularization was chosen because \emph{The Cannon} is designed for predicting large sets of elemental abundances, and it is reasonable to assume that only one or a few elemental abundances will affect the flux at a single pixel of the wavelength range. For more details on regularization or the model itself, see \citep{casey16}. 

In the training step, the log-likelihood is maximized via the Broyden-Fletcher-Goldfarb-Shanno (BFGS) algorithm to derive the best-fit model coefficients $\pmb{\theta}_{j}$ and intrinsic scatter $s_{j}^{2}$:

\begin{eqnarray}
\pmb{\theta}_{j}, s_{j}^{2} \leftarrow \underset{\pmb{\theta},s_{j}}argmax \Big [\sum_{n=0}^{N-1} \textrm{ln}p(f_{jn} | \pmb{\theta}_{j},\pmb{v}(l_{n}),s_{j}^{2})\Big] \hspace{0.5mm}.
\end{eqnarray}

\noindent Plugging in the explicit form of the log-likelihood (Equation~\ref{eq:eq4}) leads to

\begin{eqnarray}
\pmb{\theta}_{j}, s_{j}^{2} \leftarrow \underset{\pmb{\theta},s_{j}}argmax \Big [\sum_{n=0}^{N-1} - \frac{[f_{jn} -  \pmb{v}(l_{n}) \cdot \pmb{\theta}]^{2}}{\sigma_{jn}^{2} + s_{j}^{2}} \hspace{1.5mm}  \nonumber \\  - \sum_{n=0}^{N-1} \textrm{ln}(\sigma_{jn}^{2} + s_{j}^{2})  \hspace{0.5mm}-  \hspace{0.5mm} \Lambda Q(\pmb{\theta}) \Big] \hspace{0.5mm}.
\end{eqnarray}

\begin{figure*}[t]
		\centering
		\includegraphics[width=0.98\textwidth]{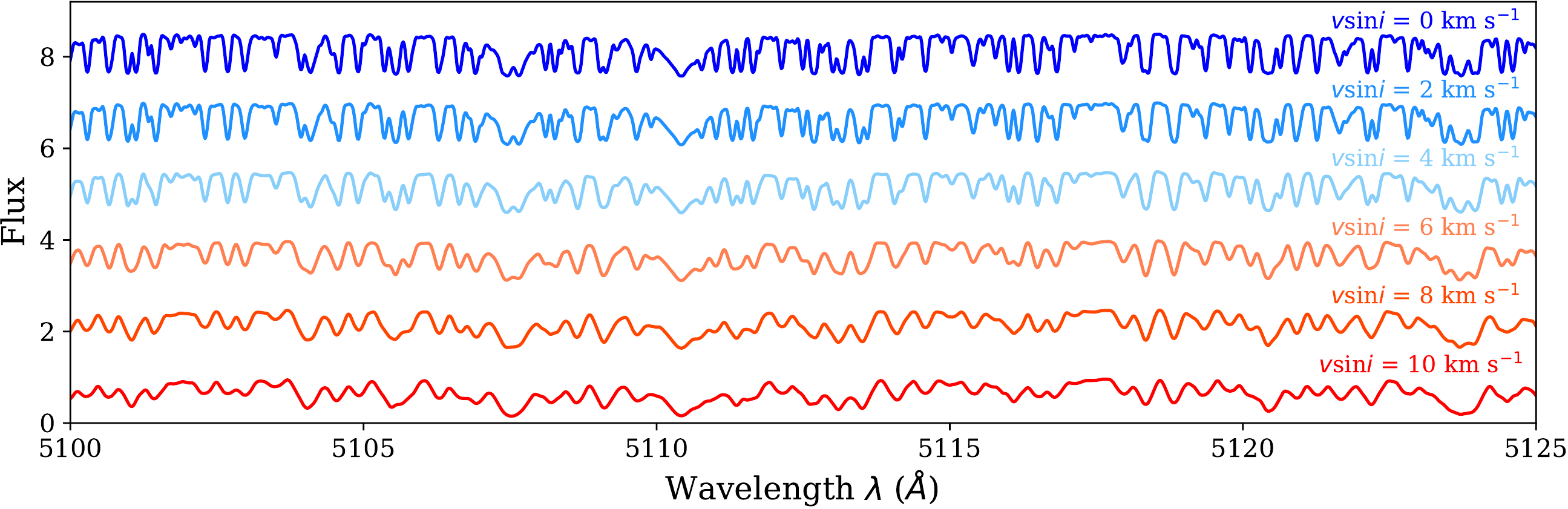}
		\caption{Synthetic spectra generated via \texttt{SpecMatch-Syn} under the same $T_{\textrm{eff}}$, $\textrm{log}g$, and [Fe/H] conditions ($T_{\textrm{eff}}$ = 4000 K, $\textrm{log}g$ = 4.5 cm s$^{-2}$, [Fe/H] = 0.2 dex) but with varying amounts of additional broadening. From top to bottom, the spectra have $v$sin$i$ = 0--10 km s$^{-1}$ in increments of +2 km s$^{-1}$.}
		\label{fig:figure3}
\end{figure*} 

\subsection{Test Step}
In the test step, we set the model labels ($\pmb{\theta}_{j}$,$s_{j}^{2}$) to the optimized values determined during the training step at every pixel $j$, and fit for the label list $l_{n}$ for each star $n$ that minimizes the log-likelihood:

\begin{eqnarray}
l_{n} \leftarrow \underset{l_{n}}argmin \Big [\sum_{j=0}^{J-1} -  \hspace{0.5mm} \frac{[f_{jn} -  \pmb{v}(l_{n}) \cdot \pmb{\theta}]^{2}}{\sigma_{jn}^{2} + s_{j}^{2}} \Big] \hspace{0.5mm},
\end{eqnarray}

\noindent Optimization of the log-likelihood in the test step is carried out via weighted least squares.

\section{\emph{The Cannon} Performance}\label{sec:cvalidation}
\subsection{Building Intuition with Synthetic Spectra}\label{sec:synthetic}
Before running \emph{The Cannon} on our cool star sample, we sought to establish a measure of baseline performance. We did this by constructing a sample of synthetic spectra that mimics the cool star sample: 141 "stars" with the same label values as the true cool sample. Because the labels of the synthetic spectra, by definition, lack uncertainty, they can provide a sense of how well \emph{The Cannon} performs under perfect conditions. The synthetic spectra are generated from the publicly-available code \texttt{SpecMatch-Syn} which fits five regions of optical spectra by interpolating within a grid of model spectra from the library described in \citet{coelho05}. For more details on \texttt{SpecMatch-Syn}, see \citet{petigura15}. See Fig. \ref{fig:figure3} for examples of synthetic spectra.

We tested the validity of \emph{Cannon}-predicted labels through a bootstrap leave-one-out cross-validation scheme where we trained the spectral model on all objects in the synthetic spectral sample but one, and predicted labels for the object that was left out. We carried out this scheme iteratively to pass through the entire sample and predict labels for every object. Following the work of \citet{ness15} and \citet{casey16}, we began with a spectral model in which the label list $l_{n}$ was quadratic in the labels, resulting in the following label list:

\begin{equation}
\begin{gathered}
l_{n} \equiv [1,T_{\textrm{eff}}, R_{*}, [Fe/H], T_{\textrm{eff}}^{2}, T_{\textrm{eff}} \cdot R_{*}, \\
T_{\textrm{eff}} \cdot [Fe/H], R_{*}^{2}, R_{*} \cdot [Fe/H], [Fe/H]^{2}] \hspace{0.5mm} ,
\end{gathered}
\end{equation}

\noindent Where 1, the first element in the label list, is there to allow for a linear offset in the fitting \citep{ness15}. We found that modeling the projected rotational velocity $v$sin$i$ as a fitted-for parameter in addition to $T_{\textrm{eff}}$, $R_{*}$, and [Fe/H] resulted in more accurate labels predictions; a second order model without $v$sin$i$ achieves accuracies of 40 K in $T_{\textrm{eff}}$, 13\% in $R_{*}$, and 0.06 dex in [Fe/H], while a second order model with $v$sin$i$ achieves accuracies of 32 K in $T_{\textrm{eff}}$, 13\% in $R_{*}$, and 0.03 dex in [Fe/H].

Using a third order rather than second order (quadratic-in-label) model with $v$sin$i$ further improves label predictions; a third order model achieves accuracies of 22 K in $T_{\textrm{eff}}$, 8\% in $R_{*}$, and 0.03 dex in [Fe/H]. 
Thus, these tests with synthetic spectra motivate a third order \emph{Cannon} model with $v$sin$i$ included as a label. The third order model results in a label list composed of additional third order cross terms, bringing the total number of terms up to 20.

\subsection{Cool Star Sample}
To run \emph{The Cannon} on the cool star sample, we employed the same bootstrap leave-one-out cross-validation scheme. As in the case of synthetic spectra, the cool star HIRES spectra are best described by a third order model, which is unsurprising given their resolution of $R$ $\approx$ 60,000 ($\sim$3 times the resolution of APOGEE spectra). The more flexible model may also better-describe our more diverse training set, composed of stars with a wider $T_{\textrm{eff}}$ range (APOGEE stars are confined to $T_{\textrm{eff}}$ = 3500--5500 K). A third order model fitting for $T_{\textrm{eff}}$, $R_{*}$, and [Fe/H] achieves precisions of 80 K in $T_{\textrm{eff}}$, 6\% in $R_{*}$, and 0.1 dex in [Fe/H]. 

We found that \emph{The Cannon} predicted anomalously poor label values for one source (GL896A). Upon inspection, we found that the spectrum of GL896A exhibits significantly broader features than any other source in our sample. Because GL896A is not well-represented in the training set, \emph{The Cannon} is unable to construct a model that well-describes GL896A (Fig. \ref{fig:figure5}, top panel). While such fast rotators are rare amongst K and M dwarfs, the presence of this target indicates that our implementation of \emph{The Cannon} must still take them into consideration. We modified our implementation by augmenting and diversifying the training set; we created $x$ copies of each spectrum in the cool star sample (exploring different values of $x$ to see which resulted in the best performance), and applied differential values of artificial broadening to simulate faster stellar rotation. Artificial broadening was carried out by convolving the spectra with a rotational-macroturbulent broadening kernel described in \citet{hirano11}. 

\begin{figure}[t]
		\centering
		\includegraphics[width=0.48\textwidth]{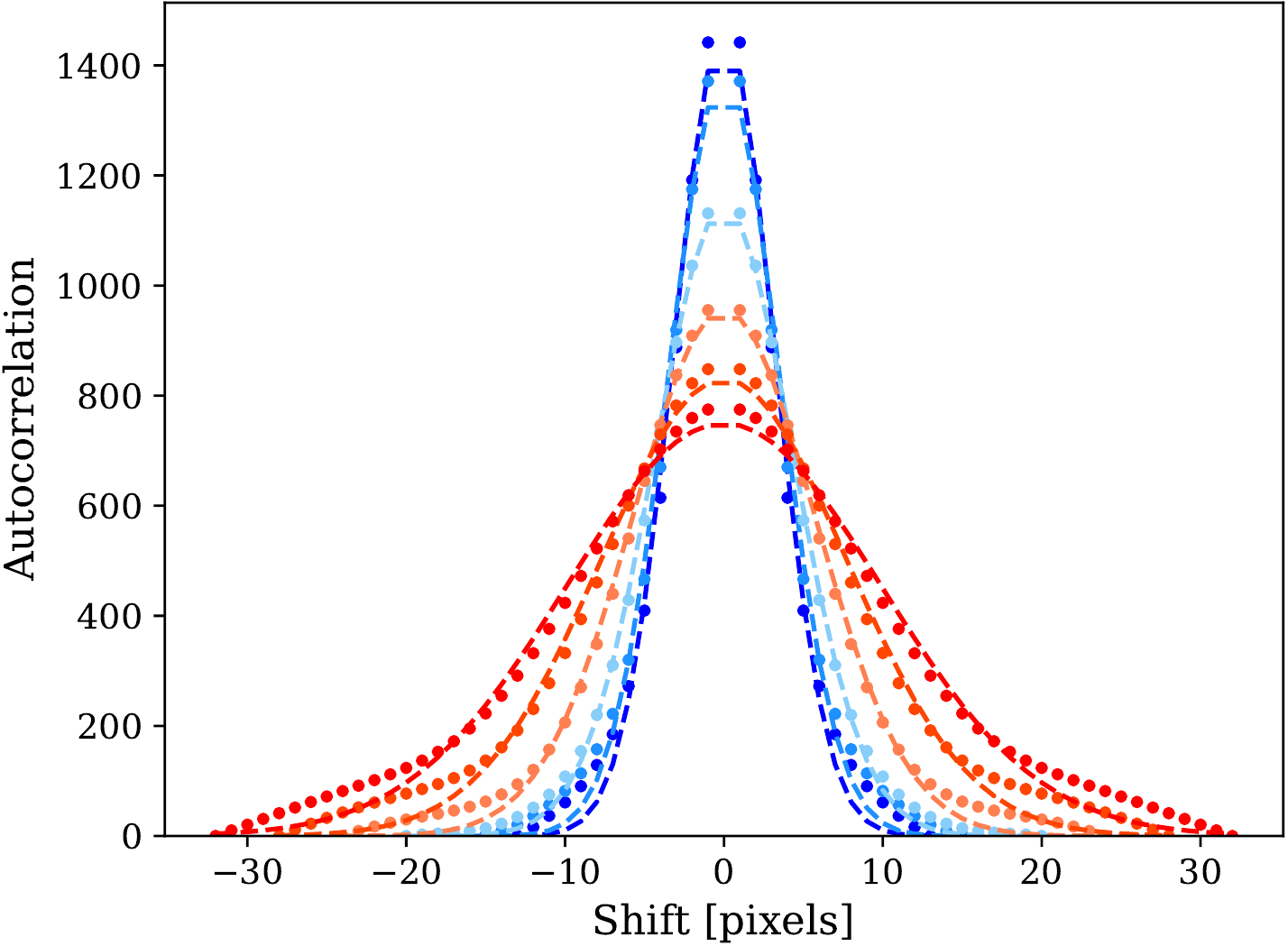}
		\caption{The autocorrelation functions of spectra displayed in Fig. \ref{fig:figure3}, marked in dotted lines. The overlaid Gaussian fits are displayed in dashed lines.}
		\label{fig:figure4}
\end{figure}

\begin{figure}[t]
		\centering
		\includegraphics[width=0.48\textwidth]{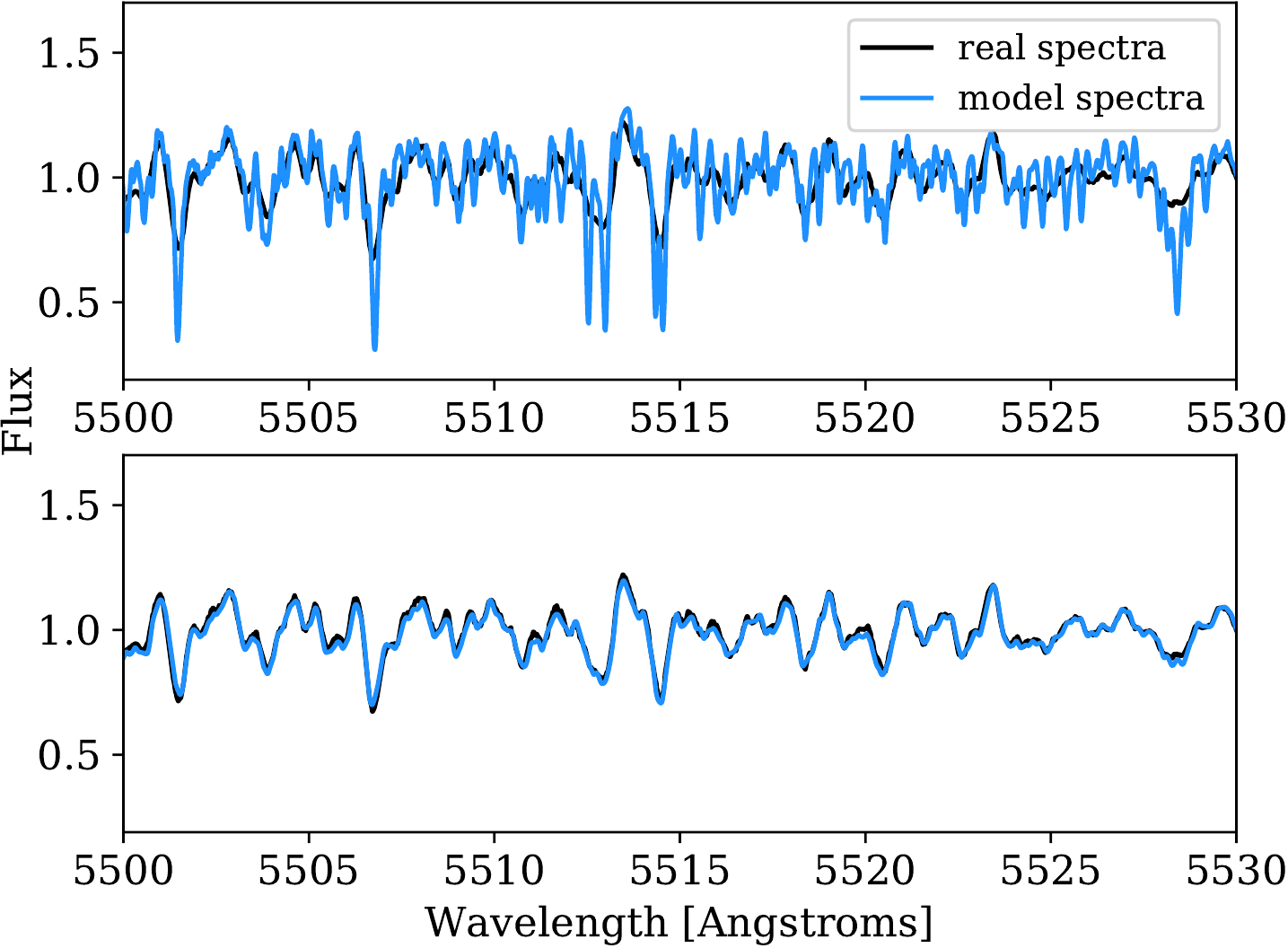}
		\caption{The spectrum of GL896A overlaid with the \emph{Cannon} model before augmenting the library with broadened copies of the spectra (top), and after (bottom). The true spectrum is plotted in black while the \emph{Cannon} models are plotted in blue.}
		\label{fig:figure5}
\end{figure}

In order for this scheme to work, $v$sin$i$ must be specified as a fitted-for label as in the tests with synthetic spectra. This is problematic because more than half of the cool stars do not have reported $v$sin$i$ values. We dealt with this by assigning all sources in the augmented sample a new label that describes general broadening, taken to be the FWHM of a Gaussian fitted to the spectral autocorrelation peaks (Fig. \ref{fig:figure4}). This resulted in better flux predictions for the spectrum of GL896A (Fig. \ref{fig:figure5}), and better label predictions overall. The most precise labels are achieved when the cool star sample is augmented by $x = 5$ (5 copies generated for each spectrum), and the copies are artificially broadened by 0--5 km s$^{-1}$ as the cool star sample does not appear to include any significantly rapid rotators ($v$sin$i$ $>$ 5 km s$^{-1}$). We ultimately achieved precisions of 68 K in $T_{\textrm{eff}}$, 5\% in $R_{*}$, and 0.08 dex in [Fe/H], and verified that these label predictions vary within the reported precisions for different \emph{Cannon} runs. While it may seem surprising that \emph{The Cannon} achieves better predictions in $R_{*}$ for empirical spectra compared to synthetic spectra (5\% versus 8\% in $R_{*}$), it should be noted that the synthetic spectra may not accurately reflect the input $R_{*}$ values as a conversion to \textrm{log}$g$ was required, which in turn required $M_{*}$. We do not have $M_{*}$ values for the cool star sample, and instead assumed a linear relationship between $R_{*}$ and  $M_{*}$ ($M_{*}/M_{\odot} = R_{*}/R_{\odot}$) to obtain mass estimates. This is a valid approximation for the main sequence, but is not perfect.

Because of the large number of terms in our model, we considered overfitting to be a potential issue. That is, overly precise modeling of the training set flux may lead to less accurate label predictions. To address this, we added regularization to our \emph{Cannon} model and assessed whether label prediction improved. We explored a grid of regularization strengths from $\Lambda = 10^{-6}$ to $\Lambda = 10^{2}$ uniform in log space. We found that no matter the regularization strength, adding regularization to the model always resulted in less precise label predictions. It is possible that regularization does not lead to better predictions for the 3 labels ($T_{\textrm{eff}}$, $R_{*}$, [Fe/H]) we are considering because all of these labels affect the flux at each wavelength point. Thus, we do not benefit from regularization that encourages sparsity (L1, encourages the model coefficients to go to zero). L1 regularization may lead to better label predictions if we expand our label set to include elemental abundances, but that is beyond the scope of this study.

\begin{figure*}[t]
    \centering
    \begin{minipage}{0.516\textwidth}
        \centering
        \includegraphics[width=0.88\textwidth]{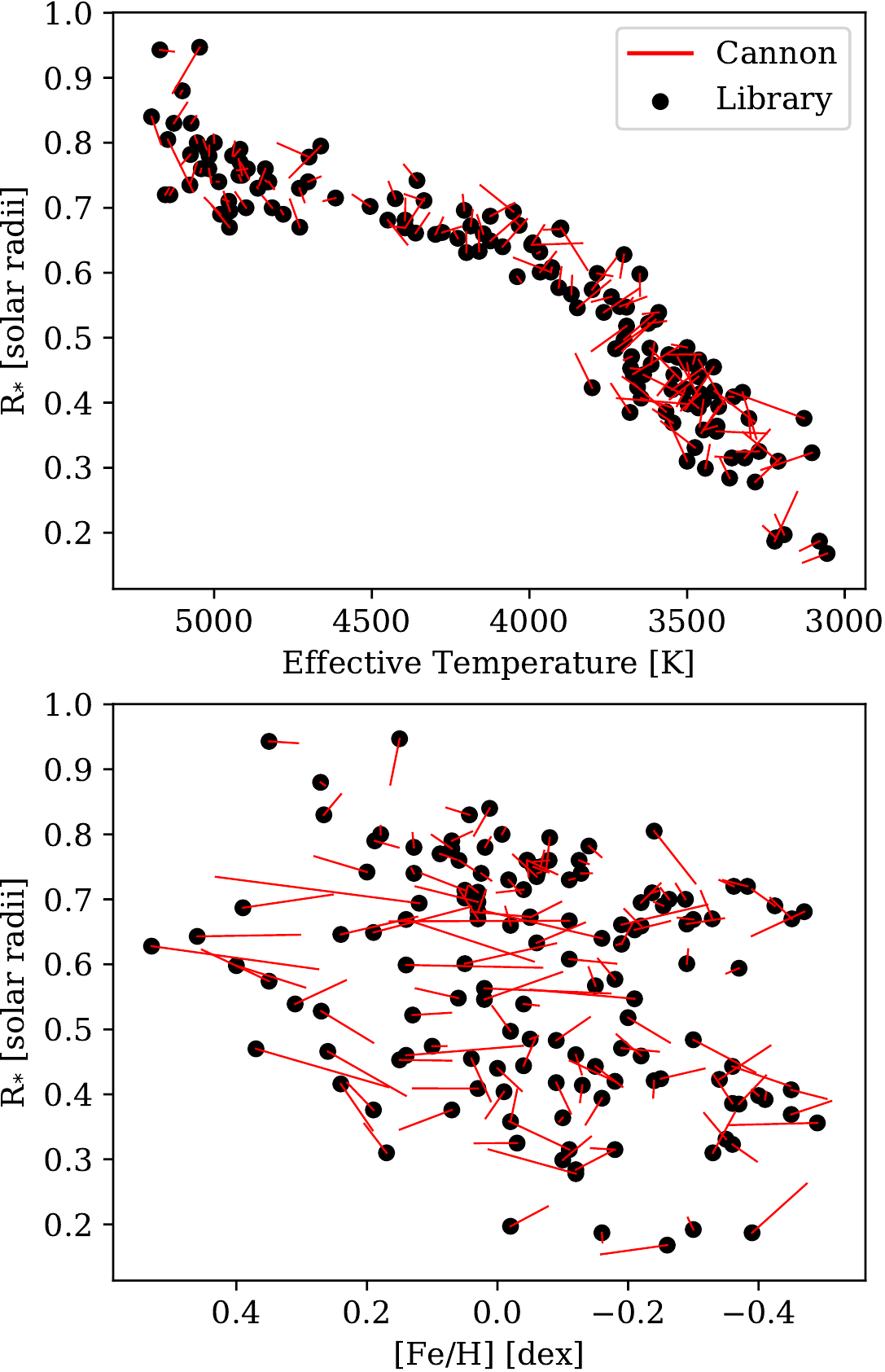} 
    \end{minipage}\hfill
    \begin{minipage}{0.48\textwidth}
        \centering
        \includegraphics[width=1.04\textwidth]{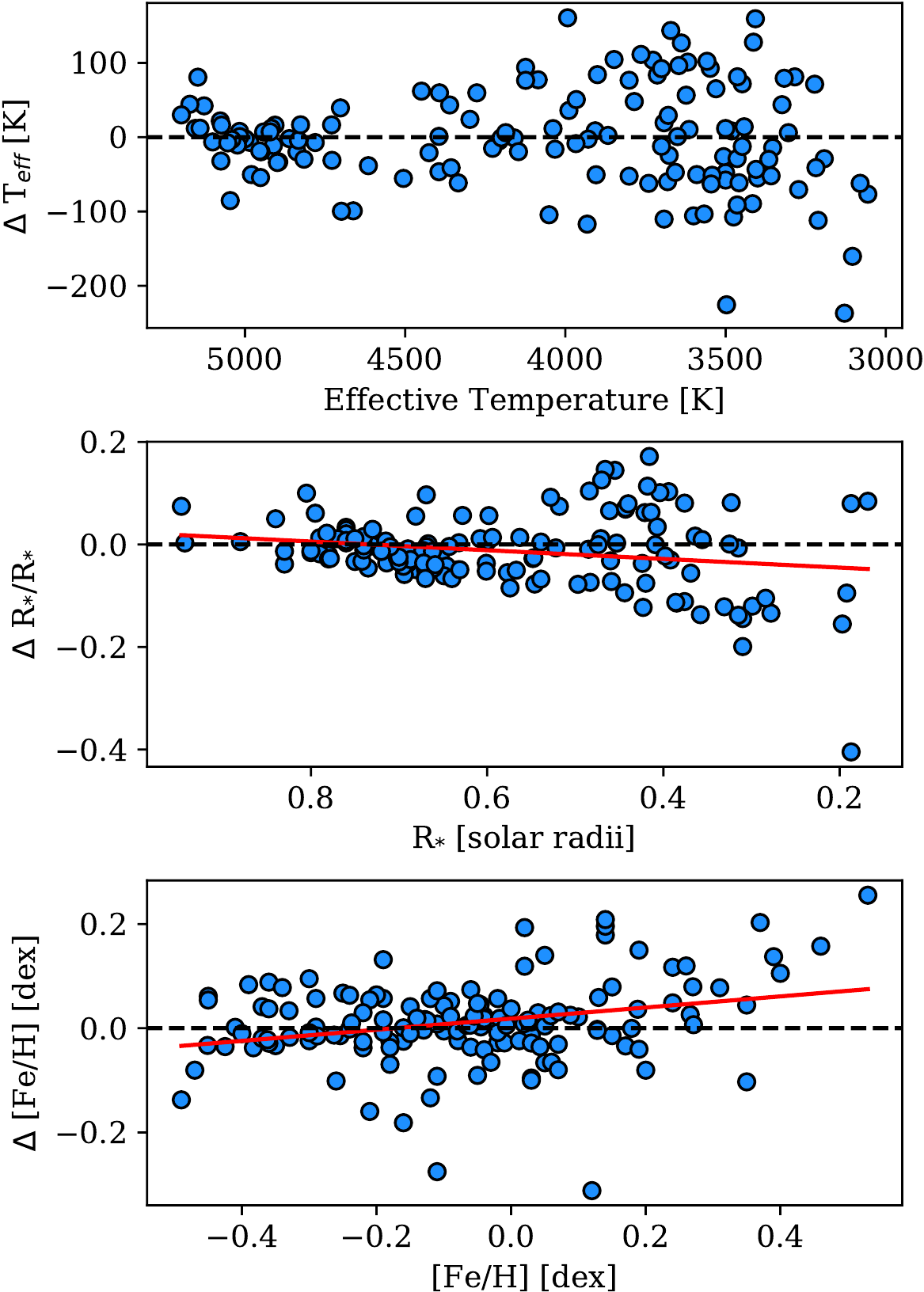} 
    \end{minipage}
    \caption{Comparison of the cool star sample library labels with the \emph{Cannon}-predicted labels ($T_{\textrm{eff}}$, $R_{*}$, [Fe/H]). In the left panel plots, the black points represent the library labels and the red lines represent the \emph{Cannon} labels. The right panel plots display the label residuals, with the red lines denoting possible trends. We note that the slope values of these linear trends are much lower than those of residuals from labels predicted via techniques that make use of empirical spectral libraries \citep{yee17}.}
    \label{fig:figure6}
\end{figure*}

\subsection{Performance at Low S/N} \label{sec:snr}
\begin{figure}[t]
		\centering
		\includegraphics[width=0.5\textwidth]{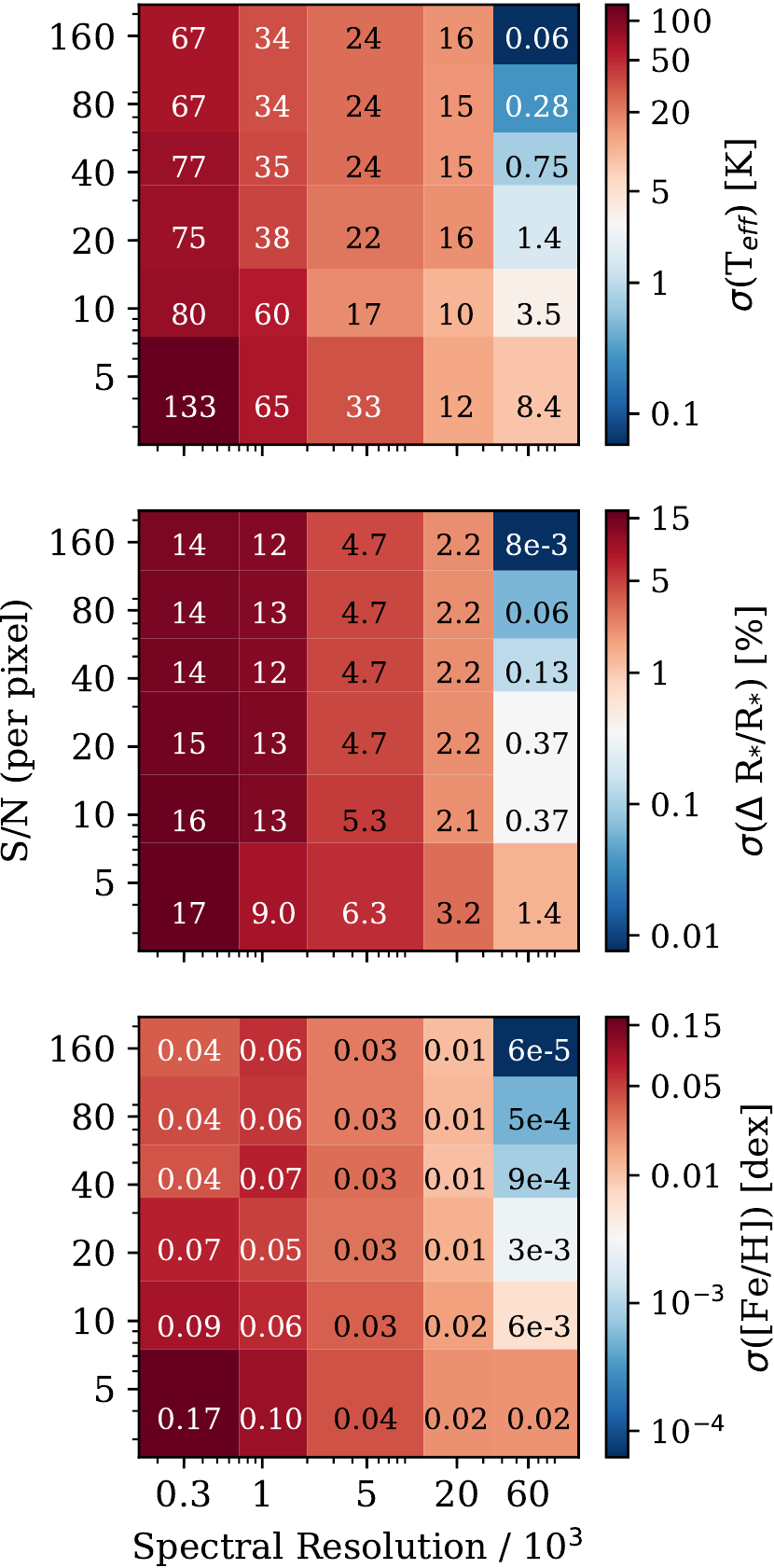}
		\caption{Log-log plots showing the median scatter of \emph{Cannon}-derived labels as a function of both S/N and resolution. Each colored block within the subplots represents the median RMS difference in $T_{\textrm{eff}}$ (top), $R_{*}$ (middle), and [Fe/H] (bottom) predictions from the cool star subset with spectra satisfying S/N $>$ 160/pixel when degraded to lower S/N and resolution. The median RMS difference is also explicitly provided within each block in units of K ($T_{\textrm{eff}}$), solar radii ($R_{*}$), and dex ([Fe/H). The median scatter increases as the S/N and resolution decreases, which is representative of the effect photon shot noise and lower resolution would have on the precision of \emph{Cannon} label predictions for HIRES spectra.}
		\label{fig:figure7}
\end{figure}

To investigate the effect of photon shot noise on the precision of label predictions made with \emph{The Cannon}, we carried out the same procedure employed by \citet{yee17} for the empirical spectroscopic tool \texttt{SpecMatch-Emp}; we isolated a subset of 20 stars from the cool star sample with S/N $>$ 160/pixel and degraded their spectra by injecting Gaussian noise to simulate target S/N values of 120, 100, 80, 60, 40, 20, and 10 per pixel. We generated 20 S/N-degraded spectra for each spectrum in the subset and S/N target value, then compared the precision of the \emph{Cannon} label predictions for the degraded spectra with those of the original S/N $>$ 160/pixel spectra, which we took as ground truth. The results are summarized in Fig. \ref{fig:figure7}.

As expected, lower S/N leads to larger median scatter in label predictions made with \emph{The Cannon}. However, the median scatter at S/N = 10/pixel is still low, with 3.5 K in $T_{\textrm{eff}}$, 0.4\% in $R_{*}$, and 0.006 dex in [Fe/H]. This demonstrates that \emph{The Cannon} is quite robust, even at low S/N values. This performance is better than that achieved by \texttt{SpecMatch-Emp}, which has median scatter values at S/N = 10/pixel of 10.4 K in $T_{\textrm{eff}}$, 1.7 \% in $R_{*}$, and 0.008 dex in [Fe/H] \citep{yee17}, though it should be noted that \texttt{SpecMatch-Emp} conducted this test with stars spanning the HR diagram while our sample is $T_{\textrm{eff}}$-limited.

Motivated by the small observed scatter in [Fe/H], we attempted to estimate the minimum change in [Fe/H] that is theoretically detectable. To do so, we considered the difference between two spectra corresponding to stars with slightly different metallicities ($\Delta$[Fe/H]). We defined a quantity $\mathcal{S}$ that relates three quantities: $\Delta$[Fe/H]; the derivative of the spectrum with changing metallicity, $\delta f / \delta [Fe/H]$; and the flux uncertainty $\sigma_{f}$.  $\mathcal{S}$ can be thought of as analogous to S/N. For the $j$th pixel, the relation is

\begin{eqnarray}
\mathcal{S}_{j} = \frac{\Big(\cfrac{\delta f}{\delta [Fe/H]}\Big )_{j} \Delta [Fe/H]}{\sigma_{f,j}} \hspace{1.5mm} .
\end{eqnarray}
This equation can be rewritten as

\begin{eqnarray}
\mathcal{S}_{j} = \frac{\Big(\cfrac{\delta f}{\delta [Fe/H]}\Big )_{j} \Delta [Fe/H]}{c_{j} \langle \sigma_{f} \rangle} \hspace{1.5mm} ,
\end{eqnarray}

\noindent where $\langle \sigma_{f} \rangle$ is the average flux uncertainty and $c_{j}$ is directly related to the blaze function. The total $\mathcal{S}$ (summing over pixels) of the metallicity measurement can be written as

\begin{eqnarray}
\mathcal{S} &=& \sqrt{\sum_{j} (\mathcal{S}_{j})^{2}} \hspace{3mm} \nonumber \\
&=& \frac{\Delta [Fe/H]}{\langle \sigma_{f} \rangle} \sqrt{\sum_{j} \Big[\Big(\frac{\delta f}{\delta [Fe/H]}\Big)/c_{j}\Big]^{2}} \hspace{1.5mm} .
\end{eqnarray}

\noindent Rearranging terms to solve for the minimum theoretically detectable change in metallicity yields

\begin{eqnarray}
\Delta [Fe/H] = \mathcal{S} \langle \sigma_{f} \rangle \bigg / \sqrt{\sum_{j} \Big[\Big(\frac{\delta f}{\delta [Fe/H]}\Big)/c_{j}\Big]^{2}} \hspace{1.5mm} .
\end{eqnarray}

\noindent A metallicity change is detectable at 1$\sigma$ if $\mathcal{S}$ = 1. For a S/N =10/pixel as considered above, i.e., $\langle \sigma_{f} \rangle = 0.1$, we find $\Delta$[Fe/H] = 0.001 dex. This is much smaller than the median scatter in [Fe/H] predictions made with \emph{The Cannon} at S/N = 10/pixel (0.006 dex). Therefore, the sensitivity of \emph{The Cannon} lies within theoretical bounds.

\subsection{Performance at Low Spectral Resolution} \label{sec:resolution}
While HIRES spectra are observed at $R$ $\approx$ 60,000, many large spectroscopic surveys are observed at lower spectral resolutions. Thus, it is valuable to quantify how spectral resolution affects the accuracies of label predictions with \emph{The Cannon}. We expected performance to decrease as spectral resolution decreases because lines will blend together, resulting in less spectral information for \emph{The Cannon} to work with. 

To investigate spectral resolution dependence, we followed the same procedure used for the S/N degradation test; we used the same subset of 20 stars with S/N $>$ 160 and simulated lower resolution by convolving their spectra with a Gaussian kernel. We again treated the label predictions of the original high resolution ($R$ $\approx$ 60,000) spectra as ground truth. We simulated spectra with target resolution values of $R$ = 50,000, 40,000, 30,000, 20,000, 10,000, 7500, and 5000. The results are summarized in Fig. \ref{fig:figure7}, which also illustrates how the precisions of label predictions are affected when both S/N and resolution are degraded.

As in the case of degraded S/N, the accuracy of \emph{Cannon} label predictions decrease with spectral resolution. At $R$ = 30,000, median scatter in the labels is 6.7 K in $T_{\textrm{eff}}$, 0.2\% in $R_{*}$, and 0.009 dex in [Fe/H]. This performance is better than that of \texttt{SpecMatch-Emp}'s at equivalent resolution, with median scatter values of 10.1 K in $T_{\textrm{eff}}$, 1.3\% in $R_{*}$, and 0.014 dex in [Fe/H] \citep{yee17}. 

\emph{The Cannon} also exhibits a much slower reduction in label accuracy as resolution continues to decrease; at $R$ = 5000, the median scatter in \emph{Cannon} predictions is 24.1 K in $T_{\textrm{eff}}$, 4.7\% in $R_{*}$, and 0.026 dex in [Fe/H], while \texttt{SpecMatch-Emp}'s is 962 K in $T_{\textrm{eff}}$, 228\% in $R_{*}$, and 0.094 dex in [Fe/H] \citep{yee17}. This suggests that \emph{The Cannon} would be a favorable method for predicting labels for spectra from many large, lower resolution spectroscopic surveys (e.g., SEGUE  \citep{beers06}, $R$ $\approx$ 2000, RAVE \citep{steinmetz06}, $R$ $\approx$ 7000, LAMOST \citep{newberg12}, $R$ $\approx$ 1800).

\subsection{Performance with Label Errors} \label{sec:label_err}
To investigate the effect of errors in the library labels on predictions made with \emph{The Cannon}, we followed the same procedure used for the S/N and resolution degradation tests; we used the same subset of 20 stars with S/N $>$ 160 and injected Gaussian noise into the labels to simulate additional uncertainty up to 1x the achievable precisions (68 K in $T_{\textrm{eff}}$, 5\% in stellar radius $R_{*}$, and 0.08 dex in [Fe/H]). We found that the labels are quite robust to realistic random noise in the library labels; adding 1x uncertainty leads to an increase in label prediction uncertainties of 22 K in $T_{\textrm{eff}}$, 4\% in $R_{*}$, and 0.06 dex in [Fe/H]. The results are summarized in Table \ref{tab:table1}.

It is worth noting that the scatter in \emph{Cannon}-predicted values of $T_{\textrm{eff}}$ is lower than the original label uncertainty by more than 50\%, suggesting that in the limit of a very large library with labels containing a certain amount of random noise, \emph{The Cannon} can derive a model that yields a higher $T_{\textrm{eff}}$ precision compared to that of the library spectra. We note that this result is insensitive to zero-point offsets; it is not possible to bootstrap to higher label precisions using \emph{The Cannon}.

\begin{deluxetable}{cccc}[h!]
\tablewidth{0.48\textwidth}
\tabletypesize{\footnotesize}
\tablewidth{0pt}
\tablecaption{Median RMS scatter in all \emph{Cannon}-derived labels after adding an 1x the amount of uncertainty to all labels \label{tab:table1}}
\tablecolumns{4}
\tablehead{
\colhead{Added uncertainty} &
\colhead{$\sigma$($T_{\textrm{eff}}$)} &
\colhead{$\sigma$($\Delta$ $R_{*}/R_{*}$)} &
\colhead{$\sigma$([Fe/H])}  \\
\colhead{($T_{\textrm{eff}}$, $R_{*}$, [Fe/H])} &
\colhead{K} &
\colhead{\%} &
\colhead{dex}
}
\startdata
68 K    & 22    & 2   & 0.02  \\
5\% $R_{*}$  & 17  & 4    & 0.02 \\
0.08 dex    & 10  & 1    & 0.06
\enddata

\end{deluxetable}

\section{Discussion} \label{sec:discussion}
We evaluated how well \emph{The Cannon}, a data-driven spectroscopic tool, is able to predict stellar labels for cool stars ($T_{\textrm{eff}}$ = 3000--5200 K) given high-resolution spectra. With adjustments to the spectral training set, it achieves precisions of 68 K in $T_{\textrm{eff}}$, 5\% in $R_{*}$, and 0.08 dex in [Fe/H]. Unlike traditional spectroscopic modeling techniques, \emph{The Cannon} does not rely on stellar models that struggle to reproduce the complexities of cool star spectra. Rather, as a data-driven method, \emph{The Cannon}'s performance improves as the input spectra become more information-rich. 

In the case of spectra with perfect labels (no uncertainty) as simulated with synthetic spectra, \emph{The Cannon} achieves label accuracies of 22 K in $T_{\textrm{eff}}$, 8\% in $R_{*}$, and 0.03 dex in [Fe/H]. \emph{The Cannon} generally makes better label predictions for synthetic spectra because the labels of real spectra include uncertainties that are endemic to the catalogs from which the cool star sample originates. These catalogs are described in \citet{vonbraun14}, \citet{mann15}, \citet{brewer16}, and \citet{yee17}, and present labels derived from a combination of modified \texttt{SME} \citep{brewer15}, photometry, parallaxes, interferometry, and empirical relations between the labels and EWs of spectral features. Each of these techniques have associated uncertainties, resulting in less precise label predictions with \emph{The Cannon} when compared to the case of spectra with perfect labels.


Compared to current synthetic spectral techniques (\texttt{SME}, \texttt{MOOG}, etc.), \emph{The Cannon} is better-suited for predicting the labels of cool stars. While the latest iterations of spectral synthesis codes model cool stars more successfully than initial versions with additions such as more accurate radiative transfer algorithms, equations of state, and larger line lists, they still lack complete sets of molecular line opacities and sufficient constraints to fully disentangle the effects of $T_{\textrm{eff}}$, \textrm{log}$g$, and abundances (e.g., \citealt{bean06, piskunov17}). 

It is more appropriate to compare \emph{The Cannon} to other data-based techniques such as \texttt{SpecMatch-Emp}, a label-predicting spectroscopic tool developed by \citet{yee17} that utilizes an empirical spectral library. While \texttt{SpecMatch-Emp} achieves accuracies of 70 K in $T_{\textrm{eff}}$, 10\% in $R_{*}$, and 0.12 dex in [Fe/H] for stars of spectral types $\sim$K4 and later, these label predictions are slightly worse than those achieved by \emph{The Cannon}. In addition, the residuals from label predictions with \texttt{SpecMatch-Emp} display linear trends where residuals are more negative for larger values in the label space, and more positive for smaller values in the label space \citep{yee17}. These trends are partly explained by considering that the empirical spectral library spans a finite region (convex hull of the label values), and is inclined to pull spectral predictions at the edge of the region towards the center. While the residuals from label predictions with \emph{The Cannon} also display slight linear trends, they are less pronounced and constitute a smaller source of systematic error (Fig. \ref{fig:figure6}, right panel). This is because the choice of flux model coefficient values allows for some extrapolation outside the finite region spanned by the training set.

While \emph{The Cannon} is a powerful tool for spectroscopic characterization, it has a number of drawbacks. For example, by individually treating each pixel within the spectral wavelength range, it assumes no covariance between flux values of any pixels. However, multiple spectral features can be affected by a single label, such as a particular elemental abundance or ionization state. This motivates converting \emph{The Cannon} into a fully Bayesian framework through the inclusion of priors such as line lists to address covariance of different spectral features, or known correlations between labels such as the Stefan-Boltzman relation. 

Although L1 regularization does not improve cool star label predictions for $T_{\textrm{eff}}$, $R_{*}$, and [Fe/H], L2 regularization may be better suited to such cases where labels do not include large sets of elemental abundances as L2 regularization does not encourage model coefficients to go to zero as rapidly. However, we are also interested in eventually using \emph{The Cannon} to predict elemental abundances, in which case L1 regularization may become a useful feature. For example, we are interested in comparing the C/O ratios of K and M dwarfs to the characteristics of planets they host as such volatile ratios can probe planet formation histories. Ultimately, we will use \emph{The Cannon} to conduct large demographic studies of cool stars with HIRES spectra with the goal of establishing correlations between small, cool stars such as K and M dwarfs and the planets they host. This work has wide potential application given that many future exoplanet surveys are focused on cool stars such as M dwarfs.

\acknowledgments

We thank Andrew Casey (Monash U.), Anna Ho (Caltech), and Melissa Ness (MPIA) for many useful discussions regarding \emph{The Cannon}. A.B. acknowledges funding from the National Science Foundation Graduate Research Fellowship under Grant No. DGE1745301. 

\newpage
\bibliography{mybib}

\begin{thebibliography}{}

\bibitem[\protect\citeauthoryear{{Bean} et~al.}{{Bean} et~al.}{2006}]{bean06}
{Bean}, J.~L., {Sneden}, C., {Hauschildt}, P.~H., {Johns-Krull}, C.~M.,  \&
  {Benedict}, G.~F. 2006, \apj, 652, 1604

\bibitem[\protect\citeauthoryear{{Beers} et~al.}{{Beers}
  et~al.}{2006}]{beers06}
{Beers}, T.~C., et~al. 2006, Memorie della Societa Astronomica Italiana, 77,
  1171

\bibitem[\protect\citeauthoryear{{Brewer} et~al.}{{Brewer}
  et~al.}{2015}]{brewer15}
{Brewer}, J.~M., {Fischer}, D.~A., {Basu}, S., {Valenti}, J.~A.,  \&
  {Piskunov}, N. 2015, \apj, 805, 126

\bibitem[\protect\citeauthoryear{{Brewer} et~al.}{{Brewer}
  et~al.}{2016}]{brewer16}
{Brewer}, J.~M., {Fischer}, D.~A., {Valenti}, J.~A.,  \& {Piskunov}, N. 2016,
  \apjs, 225, 32

\bibitem[\protect\citeauthoryear{{Casey} et~al.}{{Casey}
  et~al.}{2016}]{casey16}
{Casey}, A.~R., {Hogg}, D.~W., {Ness}, M., {Rix}, H.-W., {Ho}, A.~Q.,  \&
  {Gilmore}, G. 2016, ArXiv e-prints

\bibitem[\protect\citeauthoryear{{Coelho} et~al.}{{Coelho}
  et~al.}{2005}]{coelho05}
{Coelho}, P., {Barbuy}, B., {Mel{\'e}ndez}, J., {Schiavon}, R.~P.,  \&
  {Castilho}, B.~V. 2005, \aap, 443, 735

\bibitem[\protect\citeauthoryear{{Deen}}{{Deen}}{2013}]{deen13}
{Deen}, C.~P. 2013, \aj, 146, 51

\bibitem[\protect\citeauthoryear{{Dotter} et~al.}{{Dotter}
  et~al.}{2008}]{dotter08}
{Dotter}, A., {Chaboyer}, B., {Jevremovi{\'c}}, D., {Kostov}, V., {Baron}, E.,
  \& {Ferguson}, J.~W. 2008, \apjs, 178, 89

\bibitem[\protect\citeauthoryear{{Henry} et~al.}{{Henry}
  et~al.}{2006}]{henry06}
{Henry}, T.~J., {Jao}, W.-C., {Subasavage}, J.~P., {Beaulieu}, T.~D., {Ianna},
  P.~A., {Costa}, E.,  \& {M{\'e}ndez}, R.~A. 2006, \aj, 132, 2360

\bibitem[\protect\citeauthoryear{{Hirano} et~al.}{{Hirano}
  et~al.}{2011}]{hirano11}
{Hirano}, T., {Suto}, Y., {Winn}, J.~N., {Taruya}, A., {Narita}, N.,
  {Albrecht}, S.,  \& {Sato}, B. 2011, \apj, 742, 69

\bibitem[\protect\citeauthoryear{{Ho} et~al.}{{Ho} et~al.}{2017}]{ho17a}
{Ho}, A.~Y.~Q., et~al. 2017, \apj, 836, 5

\bibitem[\protect\citeauthoryear{{Howard} et~al.}{{Howard}
  et~al.}{2010}]{howard10}
{Howard}, A.~W., et~al. 2010, \apj, 721, 1467

\bibitem[\protect\citeauthoryear{{Mann} et~al.}{{Mann} et~al.}{2014}]{mann14}
{Mann}, A.~W., {Deacon}, N.~R., {Gaidos}, E., {Ansdell}, M., {Brewer}, J.~M.,
  {Liu}, M.~C., {Magnier}, E.~A.,  \& {Aller}, K.~M. 2014, \aj, 147, 160

\bibitem[\protect\citeauthoryear{{Mann} et~al.}{{Mann} et~al.}{2017}]{mann17}
{Mann}, A.~W., et~al. 2017, \aj, 153, 267

\bibitem[\protect\citeauthoryear{{Mann} et~al.}{{Mann} et~al.}{2015}]{mann15}
{Mann}, A.~W., {Feiden}, G.~A., {Gaidos}, E., {Boyajian}, T.,  \& {von Braun},
  K. 2015, \apj, 804, 64

\bibitem[\protect\citeauthoryear{{Ness} et~al.}{{Ness} et~al.}{2015}]{ness15}
{Ness}, M., {Hogg}, D.~W., {Rix}, H.-W., {Ho}, A.~Y.~Q.,  \& {Zasowski}, G.
  2015, \apj, 808, 16

\bibitem[\protect\citeauthoryear{{Newberg} et~al.}{{Newberg}
  et~al.}{2012}]{newberg12}
{Newberg}, H.~J., et~al. 2012, in Astronomical Society of the Pacific
  Conference Series, Vol. 458, Galactic Archaeology: Near-Field Cosmology and
  the Formation of the Milky Way, ed. W.~{Aoki}, M.~{Ishigaki}, T.~{Suda},
  T.~{Tsujimoto}, \& N.~{Arimoto}, 405

\bibitem[\protect\citeauthoryear{{Newton} et~al.}{{Newton}
  et~al.}{2014}]{newton14}
{Newton}, E.~R., {Charbonneau}, D., {Irwin}, J., {Berta-Thompson}, Z.~K.,
  {Rojas-Ayala}, B., {Covey}, K.,  \& {Lloyd}, J.~P. 2014, \aj, 147, 20

\bibitem[\protect\citeauthoryear{{Newton} et~al.}{{Newton}
  et~al.}{2015}]{newton15}
{Newton}, E.~R., {Charbonneau}, D., {Irwin}, J.,  \& {Mann}, A.~W. 2015, \apj,
  800, 85

\bibitem[\protect\citeauthoryear{{Petigura}}{{Petigura}}{2015}]{petigura15}
{Petigura}, E.~A. 2015, Ph.D. thesis, University of California, Berkeley

\bibitem[\protect\citeauthoryear{{Piskunov} \& {Valenti}}{{Piskunov} \&
  {Valenti}}{2017}]{piskunov17}
{Piskunov}, N.,  \& {Valenti}, J.~A. 2017, \aap, 597, A16

\bibitem[\protect\citeauthoryear{{Rojas-Ayala} et~al.}{{Rojas-Ayala}
  et~al.}{2012}]{rojas_ayala12}
{Rojas-Ayala}, B., {Covey}, K.~R., {Muirhead}, P.~S.,  \& {Lloyd}, J.~P. 2012,
  \apj, 748, 93

\bibitem[\protect\citeauthoryear{{Sneden}}{{Sneden}}{1973}]{sneden73}
{Sneden}, C.~A. 1973, Ph.D. thesis, THE UNIVERSITY OF TEXAS AT AUSTIN.

\bibitem[\protect\citeauthoryear{{Steinmetz} et~al.}{{Steinmetz}
  et~al.}{2006}]{steinmetz06}
{Steinmetz}, M., et~al. 2006, \aj, 132, 1645

\bibitem[\protect\citeauthoryear{{Valenti} et~al.}{{Valenti}
  et~al.}{2009}]{valenti09}
{Valenti}, J.~A., et~al. 2009, \apj, 702, 989

\bibitem[\protect\citeauthoryear{{Valenti} \& {Fischer}}{{Valenti} \&
  {Fischer}}{2005}]{valenti05}
{Valenti}, J.~A.,  \& {Fischer}, D.~A. 2005, \apjs, 159, 141

\bibitem[\protect\citeauthoryear{{Valenti} \& {Piskunov}}{{Valenti} \&
  {Piskunov}}{1996}]{valenti96}
{Valenti}, J.~A.,  \& {Piskunov}, N. 1996, \aaps, 118, 595

\bibitem[\protect\citeauthoryear{{Vogt} et~al.}{{Vogt} et~al.}{1994}]{vogt94}
{Vogt}, S.~S., et~al. 1994, in \procspie, Vol. 2198, Instrumentation in
  Astronomy VIII, ed. D.~L. {Crawford} \& E.~R. {Craine}, 362

\bibitem[\protect\citeauthoryear{{von Braun} et~al.}{{von Braun}
  et~al.}{2014}]{vonbraun14}
{von Braun}, K., et~al. 2014, \mnras, 438, 2413

\bibitem[\protect\citeauthoryear{{Yee}, {Petigura}, \& {von Braun}}{{Yee}
  et~al.}{2017}]{yee17}
{Yee}, S.~W., {Petigura}, E.~A.,  \& {von Braun}, K. 2017, \apj, 836, 77

\end{thebibliography}



\end{document}